\documentclass[aps,showpacs]{revtex4}
\usepackage{epsfig,amssymb,bm,graphics,color}
\newcommand{\eq}{\begin{eqnarray}}
\newcommand{\en}{\end{eqnarray}}

\begin{document}{\normalsize }

\title{Can parameters of $f_0$-mesons be determined correctly
analyzing only $\pi\pi$ scattering?}

\author{Yu.S. Surovtsev$^1$, P.~Byd\v{z}ovsk\'y$^2$,
R.~Kami\'nski$^3$, V.E. Lyubovitskij$^4$, M.~Nagy$^5$
\vspace*{0.25\baselineskip}}
\affiliation{ $^1$
Bogoliubov Laboratory of Theoretical Physics, Joint
Institute for Nuclear Research, 141 980 Dubna, Russia\\
$^2$ Nuclear Physics Institute, Czech Academy of Sciences,
\v{R}e\v{z} near Prague 25068,
Czech Republic\\
$^3$ Institute of Nuclear Physics, Polish Academy of Sciences,
Cracow 31342, Poland\\
$^4$ Institut f\"ur Theoretische Physik, Universit\"at T\"ubingen, 
Kepler Center for Astro and Particle Physics, \\
Auf der Morgenstelle 14, D-72076 T\"ubingen, Germany\\
$^5$ Institute of Physics, Slovak Academy of Sciences, Bratislava
84511, Slovak Republic}

\begin{abstract}
The coupled processes -- the $\pi\pi$ scattering and $\pi\pi\to K\overline{K}$ in
the $I^GJ^{PC}=0^+0^{++}$ channel -- are analyzed (both separately and combined)
in a model-independent approach based on analyticity and unitarity and using a
uniformization procedure.
It is shown: 1) a structure of the Riemann surface of the $S$-matrix for
considered coupled processes must be allowed for calculating both amplitudes and
resonance parameters, such as the mass and width;
2)~the combined analysis of coupled processes is needed as the analysis
of only $\pi\pi$ channel does not give correct values of resonance parameters even
if the Riemann surface structure is included.
\end{abstract}


\maketitle

The study of scalar mesons is very important for such profound topics in
particle physics as the QCD vacuum. However, despite of a big effort devoted
to studying various aspects of the problem~\cite{PDG12} (for recent reviews
see \cite{Amsler04,Bugg04,Close02,Klempt07}) a description of this sector
is far from being complete. Parameters of the scalar mesons, their nature and
status of some of them are still not well settled~\cite{PDG12}. E.g., applying
our model-independent method in three-channel analyses of processes
$\pi\pi\!\to\!\pi\pi,K\overline{K},\eta\eta (\eta\eta^\prime)$
\cite{SBKN-PRD10,SBL-arXiv11} we have obtained parameters of the $f_0(600)$
and $f_0(1500)$ which differ considerably from results of analyses utilizing
other methods (mainly those based on the dispersion relations and Breit--Wigner
approaches). Reasons of this difference should be understood because our
method is based only on a demand for analyticity and unitarity of the
amplitudes using a uniformization procedure. The construction of amplitudes
is practically free from any dynamical (model) assumptions using only the
{\it mathematical} fact that a local behavior of analytic functions determined
on the Riemann surface is governed by the nearest singularities on all
corresponding sheets. I.e., the obtained parameters of resonances can be considered
as free from theoretical prejudice.

First note that in our previous three-channel analyses with the uniformizing
variables \cite{SBKN-PRD10,SBL-arXiv11} we were enforced to construct a
four-sheeted model of the eight-sheeted Riemann surface. This we have achieved
neglecting the $\pi\pi$-threshold branch-point which means that we have
considered the nearest to the physical region semi-sheets of the initial
Riemann surface. This is in the line with our approach of a consistent account
of the nearest singularities on all relevant sheets. The two-channel analysis
utilizes the full Riemann surface and is, therefore, free of these approximations.
To verify a plausibility of our assumptions in the three-channel calculations,
we have performed a combined two-channel analysis of data on
$\pi\pi\to\pi\pi,K\overline{K}$ to check
whether the results of our three-channel analyses \cite{SBKN-PRD10,SBL-arXiv11}
are also obtained in the two-channel consideration. Moreover, to better
understand reasons for the above-indicated difference in results, we have
performed first the analysis only of the $\pi\pi$ scattering data in the
two-channel approach.

The two-channel $S$-matrix is determined on the four-sheeted Riemann surface.
The matrix elements $S_{ij}$, where $i,j=1(\pi\pi),2(K\overline{K})$ denote
channels, have the right-hand cuts along the real axis of the $s$ complex
plane ($s$ is the invariant total energy squared), starting with the channel
thresholds $s_i$, and the left-hand cuts related to the crossed channels.
The Riemann-surface sheets are numbered according to the signs of
analytic continuations of the roots $\sqrt{s-s_i}$ as follows:
$\mbox{signs}\bigl(\mbox{Im}\sqrt{s-s_1},\mbox{Im}\sqrt{s-s_2}\bigl)=
++,-+,--,+-$ correspond to sheets I, II, III, IV.

The resonance representations on the Riemann surface are obtained from
formulas \cite{KMS96} which express analytic continuations of the $S$-matrix
elements to unphysical sheets in terms of those on the physical (I) sheet
having only resonance zeros (beyond the real axis). Then, starting from the
resonance zeros on sheet I, one can obtain an arrangement of poles and zeros
of a resonance on the whole Riemann surface (``pole clusters''). In the
two-channel case, according to these formulas and a real analyticity
amplitudes of all coupled processes have conjugate poles at the same
points of complex energy on sheets II, III and IV if $S_{11}^{\rm I}=0$,
$S_{11}^{\rm I}S_{22}^{\rm I}-(S_{12}^{\rm I})^2=0$ and $S_{22}^{\rm I}=0$,
respectively (the superscript I means the matrix elements on sheet I).
Therefore three types of pole clusters representing states of different
nature arise: ({\bf a}) when there is a pair of conjugate zeros only in
$S_{11}^{\rm I}=0$, ({\bf b}) only in $S_{22}^{\rm I}=0$, and ({\bf c}) in
both the $S_{11}^{\rm I}=0$ and $S_{22}^{\rm I}=0$. If the coupling of channels
is present ($S_{12}\not=0$) (i.e. a state decays into both channels and/or
is exchanged in crossing channels), then this state being of type {\bf a}
is represented in $S_{11}$ by a pair of conjugate poles on sheet~II and
a pair of conjugate zeros on sheet~I and also by a pair of
conjugate poles on sheet~III and a pair of conjugate zeros on sheet~IV
at the same complex-energy points, which are to be shifted with respect
to the zeros on sheet~I. For states of type {\bf b}, the pair of conjugate
poles on sheet~III is shifted relative to the pair of poles on sheet~IV. For
the states of type {\bf c}, one must consider two pairs of conjugate poles
on sheet~III. Generally, wide multi-channel states are most adequately
represented by
pole clusters, because the pole positions are rather stable characteristics
for various models, whereas masses and widths are very model-dependent
\cite{SKN-epja02}. For calculation of the latters one must use the poles
on those sheets where they are not
shifted (due to the channel couplings) with respect to the zero position
on sheet I. For resonances of types {\bf a} and {\bf b} these poles
are on sheets~II and IV, respectively. For resonance of type {\bf c}
the poles can be used on both these sheets.
In the case of N channels, the poles only on the sheets with the numbers
$2^i$ ($i=1,\cdots,N$ is the number of channel), i.e. II, IV, VIII,$\ldots$,
should be used for calculating resonance parameters \cite{SBL-arXiv11,KMS96}.

It is convenient to use the Le Couteur--Newton relations~\cite{LeCou}.
They express the $S$-matrix elements of all coupled processes in terms
of the Jost matrix determinant $d(\sqrt{s-s_1},\cdots,\sqrt{s-s_N})$
that is a real analytic function with the only square-root branch-points at
$\sqrt{s-s_\alpha}=0$.

A necessary and sufficient condition for existence of the multi-channel
resonance is its representation by one of the types of pole clusters.
To use this representation of resonances, which is very important
for the wide multi-channel states, a uniformizing variable is applied.
Analyzing $\pi\pi\to\pi\pi,K\overline{K}$ we applied the uniformizing
variable \cite{SKN-epja02} which takes into account, in addition to the
$\pi\pi$- and $K\overline{K}$-threshold branch-points, the left-hand
branch-point at $s\!=0$, related to the $\pi\pi$ crossed channels:
\begin{equation}\label{lv}
v=\frac{m_K\sqrt{s-4m_\pi^2}+m_\pi\sqrt{s-4m_K^2}}{\sqrt{s(m_K^2-m_\pi^2)}}.
\end{equation}
It maps the four-sheeted Riemann surface with two unitary cuts and the
left-hand cut onto the $v$-plane.
Representation of resonances of various types on the uniformization $v$-plane
can be found in~\cite{SBKLN-1206}.

On the $v$-plane, $S_{11}(v)$ has no cuts; $S_{12}^2(v)$ and $S_{22}(v)$
do have the cuts which arise from the left-hand cut on the $s$-plane,
starting at $s=4(m_K^2-m_\pi^2)$, which is further approximated by a pole
$$d_L=v^{-4}\Bigl(1-\bigl(p-i\sqrt{1-p^2}\bigr)v\Bigr)^4\Bigl(1+
\bigl(p+i\sqrt{1-p^2}\bigr)v\Bigr)^4\,,$$
where $p=0.903\pm0.0004$ from analysis. An explanation
of the fourth power of this pole can be found in \cite{SBKLN-1206}.

On the $v$-plane, the function $d(v)$ in the Le Couteur--Newton relations
\cite{LeCou,SKN-epja02} does not possess any branch points.
The main model-independent contribution of resonances, given by the pole
clusters, is factorized in the $S$-matrix elements from the background.
The possible remaining small (model-dependent) contributions of resonances
are supposed to be included in the background. Therefore, $d(v)$ is taken
as $d=d_{res}d_Ld_{bg}$ where the resonance part is
\begin{equation}
d_{res} = v^{-M}\prod_{n=1}^{M} (1-v_n^* v)(1+v_n v)
\end{equation}
with $M$ the number of pairs of the conjugate zeros.
The background part is
\begin{equation}
d_{bg}=\mbox{exp}\Bigl[-i\sum_{n=1}^{3}(\sqrt{s-s_n}/2m_n)(\alpha_n+i\beta_n)
\Bigr],
\end{equation}
$$
\alpha_n,\beta_n=a_{n1},b_{n1}+ \sum_{k=\eta,\sigma,v}a_{nk},b_{nk}(s/s_k-1)\theta(s-s_k)
$$
with $s_\eta$ and $s_\sigma$ the $\eta\eta$ and $\sigma\sigma$
thresholds, respectively, $s_v$ a combined threshold of the
$\eta\eta^{\prime}$, $\rho\rho$ and $\omega\omega$ thresholds; from
the analysis: $s_\sigma=1.6558~{\rm GeV}^2$, $s_v=2.1293~{\rm
GeV}^2$.

At present in the scalar sector, there are alternative data for the $\pi\pi$
scattering -- \cite{Hya73} and \cite{Kamin02} -- which are different considerably
in the 0.76~GeV region and especially above 1.45~GeV. Therefore separate analyses
using these alternative data are needed. Here we performed the analysis taking
for the $\pi\pi$ scattering in interval $0.575~{\rm GeV}\!<\!\sqrt{s}\!< 1.89~{\rm GeV}$
the data from \cite{Hya73} and for $\sqrt{s} < 1~{\rm GeV}$ from many works.
References to the latter and to practically all accessible $\pi\pi\to K\overline{K}$
data used can be found in \cite{SBKLN-1206}.
Analysis using the data \cite{Kamin02} will be presented in other paper.

First analyzing only the $\pi\pi$ scattering, we supposed an existence
of two states (narrow $f_0(1500)$ and wide $f_0^\prime(1500)$) and
achieved an excellent description for the phase shift $\delta_{11}$ and
modulus $|S_{11}|$ (the total $\chi^2/\mbox{NDF}\!\approx\!1.07$) with
the resonance parameters (Table~\ref{tab:pole_clusters_pipi}) which largely
coincide with estimations of the PDG \cite{PDG12} (cf. also \cite{GarciaMKPRE-11}).
The only distinction is observation of the wide $f_0^\prime(1500)$. The fact
that this state is not observed in works cited by the PDG is related, it
seems, with peculiarities of analysis of data therein.

\begin{table}[htb!]
\caption{The pole positions of resonances 
$\sqrt{s_r}\!=\!{\rm E}_r\!-\!i\Gamma_r/2$ [MeV] on sheets~II and IV in 
the $\sqrt{s}$-plane in the analysis of only $\pi\pi$-scattering.}
\label{tab:pole_clusters_pipi}
\begin{center}
{
\vskip-0.2truecm
\hspace*{-0.32cm}
\def\arraystretch{.65}
\begin{tabular}{ccc}
\hline\hline
 {Sheet} & II & IV \\
\hline
{$f_0(600)$} & 447.5$\pm$5.9--i(267$\pm$6.5) & {} \\
{$f_0(980)$} & 1001.1$\pm$3.7--i(20.3$\pm$2.6) & {} \\
{$f_0(1370)$}& {} & 1301.1$\pm$47.9--i(224$\pm$49.3) \\
{$f_0(1500)$}& {} & 1503.7$\pm$45.1--i(56.5$\pm$39.4) \\
{$f_0^\prime(1500)$} & 1511.4$\pm$11.2--i(200.5$\pm$11) &
1505.9$\pm$38.5--i(168$\pm$40.6) \\
{$f_0(1710)$}& {} & 1720$\pm$32.2--i(64.9$\pm$30.1) \\
\hline\hline
\end{tabular}}
\end{center}
\end{table}

\vspace*{-.5cm}
In the analysis, the $f_0 (600)$ and $f_0 (980)$ are described by
the clusters of type {\bf a}; $f_0 (1370)$, $f_0(1500)$ and
$f_0(1710)$, type {\bf b}; $f_0^\prime(1500)$, type {\bf c}. The
received background parameters are:
$a_{11}\!=\!-0.0895\!\pm\!0.0030$, $a_{1\eta}\!=\!0.04\!\pm\!0.03$,
$a_{1\sigma}\!=\!0.0\!\pm\!0.8$, $a_{1v}\!=\!0.0\!\pm\!0.7$,
$b_{11}\!=\!0.0\!\pm\!0.007$, $b_{1\eta}\!=\!0.0\!\pm\!0.01$,
$b_{1\sigma}\!=\!0.0\!\pm\!0.02$, $b_{1v}\!=\!0.054\!\pm\!0.036$.

\begin{figure}[!htb]
\begin{center}
\includegraphics[width=0.48\textwidth,angle=0]{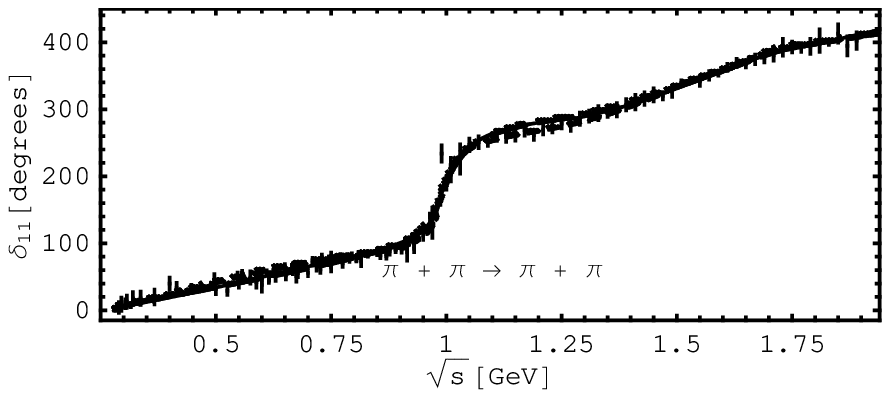}
\includegraphics[width=0.48\textwidth,angle=0]{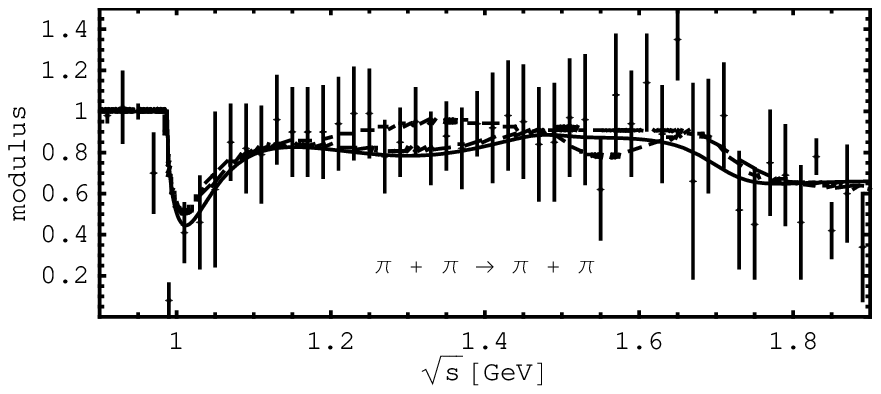}\\
\includegraphics[width=0.48\textwidth,angle=0]{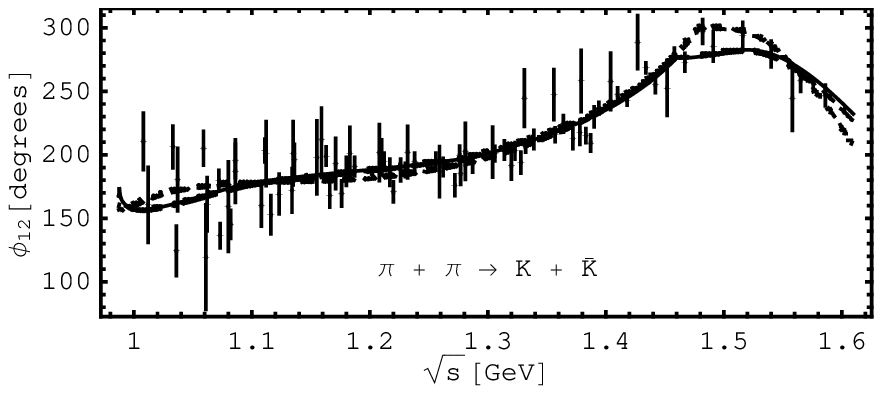}
\includegraphics[width=0.48\textwidth,angle=0]{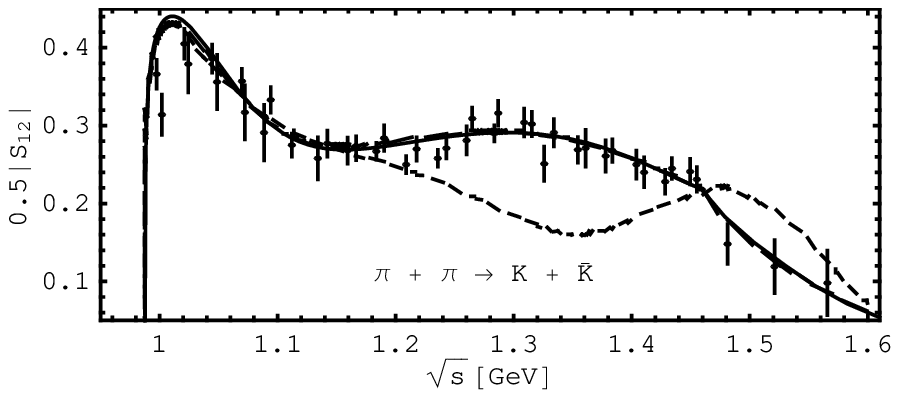}
\caption{The $S$-wave phase shifts and modules of the $\pi\pi$-scattering
and $\pi\pi\to K\overline{K}$ matrix elements.
The short-dashed lines correspond to the analysis only of the
$\pi\pi$ scattering.
The long-dashed and solid lines correspond to solutions A and B of the
combined analysis of $\pi\pi\to\pi\pi,K\overline{K}$, respectively.
The data are from
Refs.~\cite{Hya73,SBKLN-1206,Rosselet77,Bel'kov79}.}
\label{fig:set_I}
\end{center}
\end{figure}

In Fig.~\ref{fig:set_I}, we show the fitting only to the $\pi\pi$ scattering data
and energy behavior of the phase shift $\phi_{12}$ and modulus of $S_{12}$
calculated using the resonance parameters from this analysis.
In spite of the very good description of data and the very good agreement
of obtained $\pi\pi$ scattering length $a_0^0$ with the experimental
results and with the chiral perturbation theory (ChPT) calculations
(see Table~\ref{tab:scattering_length}), this analysis shows two important
flaws:
{\bf (1)} The negative background phase-shift beginning at the $\pi\pi$
threshold ($a_{11}\!=\!-0.0895$) is necessary for a successful description
of the data. This should not be the case because, in the uniformizing variable,
we have allowed for the left-hand branch-point at $s\!=0$ which gives a main
contribution to the $\pi\pi$ background below 1~GeV.
Other possible contributions of the left-hand cut from exchanges by
the nearest $\rho$ and $f_0 (600)$ mesons practically
obliterate each other \cite{SKN-epja02} because vector and scalar particles
contribute with the opposite signs due to gauge invariance.
{\bf (2)} Description of the $\pi\pi\to K\overline{K}$ data, using the same
parameters of resonances as in the $\pi\pi$ channel, is satisfactory only for
the phase shift $\phi_{12}$ which is due to the approximation of the left-hand
cut in $S_{12}$ and $S_{22}$ by the fourth-power pole. The
modulus $|S_{12}|$ is described well only from the $K\overline{K}$ threshold
to about 1.15~GeV as it should be due to the two-channel unitarity. Above
this energy the description fails even qualitatively (Fig.~\ref{fig:set_I}).

From this we conclude: If the data are consistent, for obtaining correct
parameters of wide resonances the combined analysis of coupled processes is
needed. Further that analysis of $\pi\pi\to\pi\pi,K\overline{K}$ is
performed successfully. The data for the $\pi\pi$ scattering below 1~GeV admit
two solutions for the phase shift -- A and B -- which differ mainly in the
pole position on sheet~II of the $f_0(600)$. The total $\chi^2/\mbox{NDF}$ is
1.53 for the A-solution and 1.44 for B-solution.
The resonances are described by pole clusters of the same types as in the
analysis only of the $\pi\pi$ scattering. In Table~\ref{tab:pole_clusters_A_I}
we show the pole positions of resonances on sheets~II and IV on the
$\sqrt{s}$-plane.

\begin{table}[htb!]
\caption{The pole positions of resonances on sheets~II and IV in the $\sqrt{s}$-plane
in the combined analysis of the $\pi\pi\to\pi\pi,K\overline{K}$ data. The complete
pole-clusters of resonances can be found in~\cite{SBKLN-1206}.}
\label{tab:pole_clusters_A_I}
\begin{center}
{
\hspace*{-0.5cm}
\def\arraystretch{.65}
\begin{tabular}{ccc}
\hline\hline
Sheet & II & IV \\
\hline & \multicolumn{2}{c}{A-solution}\\ \hline
$f_0(600)$ & 517$\pm$7.8--i(393.9$\pm$6) & {} \\
$f_0(980)$ & 1004.6$\pm$3.9--i(25.0$\pm$2.3) & {} \\
$f_0(1370)$ &  & 1342.9$\pm$12.2--i(221.6$\pm$30.7) \\
$f_0(1500)$ & {} & 1501.1$\pm$6.4--i(56.6$\pm$6.0) \\
$f_0^\prime(1500)$ & 1532.2$\pm$12.4--i(323.2$\pm$21) &
1519.3$\pm$18.7--i(339.5$\pm$42.2) \\
$f_0 (1710)$ & {} & 1717$\pm$34.9--i(72.9$\pm$16.2) \\
\hline & \multicolumn{2}{c}{B-solution}\\
\hline $f_0(600)$ & 550.6$\pm$9--i(502.1$\pm$7.2) & {} \\
$f_0(980)$ & 1003.2$\pm$3--i(28.9$\pm$2) & {} \\
$f_0(1370)$ & {} & 1336.7$\pm$14--i(251.9$\pm$27.5) \\
$f_0(1500)$ & {} & 1500.3$\pm$6.3--i(57.0$\pm$6.4) \\
$f_0^\prime(1500)$ & 1528.4$\pm$12.5--i(328$\pm$20.2) &
1515.6$\pm$17--i(340.3$\pm$34.9) \\
$f_0 (1710)$ & {} & 1722$\pm$35.7--i(92.3$\pm$20.3) \\
\hline\hline
\end{tabular}}
\end{center}
\end{table}
The obtained background parameters for the A-so\-lu\-ti\-on are:
$a_{11}\!=\!0.0\!\pm\!0.003$, $a_{1\eta}\!=\!-0.1004\!\pm\!0.0301$,
$a_{1\sigma}\!=\!0.2148\!\pm\!0.0822$, $a_{1v}\!=\!0.0\!\pm\!0.07$,
$b_{11}\!=\!b_{1\eta}\!=\!b_{1\sigma}\!=\!0$, $b_{1v}\!=\!0.012\!\pm\!0.0287$,
$a_{21}\!=\!-0.919\!\pm\!0.107$, $a_{2\eta}\!=\!-1.399\!\pm\!0.348$,
$a_{2\sigma}\!=\!0.0\!\pm\!0.7$, $a_{2v}\!=\!-11.45\!\pm\!0.75$,
$b_{21}\!=\!0.0747\!\pm\!0.0503$, $b_{2\eta}\!=\!b_{2\sigma}\!=\!0$,
$b_{2v}\!=\!4.83\!\pm\!1.94$;
for B-solution: $a_{11}\!=\!0.0\!\pm\!0.003$, $a_{1\eta}\!=\!-0.0913\!\pm\!0.0327$,
$a_{1\sigma}\!=\!0.1707\!\pm\!0.0899$, $a_{1v}\!=\!0.0\!\pm\!0.07$,
$b_{11}\!=\!b_{1\eta}\!=\!b_{1\sigma}\!=\!0$, $b_{1v}\!=\!0.006\!\pm\!0.029$,
$a_{21}\!=\!-1.338\!\pm\!0.111$, $a_{2\eta}\!=\!-1.119\!\pm\!0.376$,
$a_{2\sigma}\!=\!0.0\!\pm\!0.8$, $a_{2v}\!=\!-12.13\!\pm\!0.77$,
$b_{21}\!=\!0.018\!\pm\!0.050$, $b_{2\eta}\!=\!b_{2\sigma}\!=\!0$,
$b_{2v}\!=\!4.48\!\pm\!1.98$.

In the combined analysis both flaws of the only
$\pi\pi$-scattering analysis are cured. Now the $\pi\pi$ background below
the $K\overline{K}$ threshold is absent ($a_{11}\!=\!0.0$). An arising
pseudo-background at the $\eta\eta$ threshold ($a_{1\eta}\!<\!0$) is also
clear: this is a direct indication
to consider explicitly the $\eta\eta$-threshold branch-point. This was
already done in our work \cite{SBL-arXiv11}. In the combined
analysis the $f_0(600)$ parameters are changed considerably receiving
new values closer to those obtained in our three-channel analysis
\cite{SBL-arXiv11}. Earlier one noted that wide resonance parameters
are largely controlled by the non-resonant background \cite{Achasov-Shest}.
In part this problem is removed due to allowing
for the left-hand branch-point at $s=0$ in the uniformizing variable.

In the Table~\ref{tab:scattering_length} we compare our results for the $\pi\pi$
scattering length $a_0^0$ with results of some other theoretical and
experimental works.

\begin{table}[!htb!]
\caption{The $\pi\pi$ scattering length $a_0^0$.}
\label{tab:scattering_length}
\begin{center}
{
\def\arraystretch{.7}
\hspace*{-.3cm}
\begin{tabular}{llc} \hline\hline $a_0^0$
[$m_{\pi^+}^{-1}$] & Remarks & References
\\
\hline $0.222\pm 0.008$ & Analysis only of  & This work \\
 & $\pi\pi$ scattering & \\
$0.230\pm 0.004$ & A-solution & This work \\
$0.282\pm 0.003$ & B-solution & This work  \\
$0.26\pm 0.05$ & Analysis of the $K\to\pi\pi e\nu$ &
\cite{Rosselet77} \\ {} & using Roy's equation &
{}
\\
$0.24\pm 0.09$ & Analysis of $\pi^-p\to\pi^+\pi^-n$ & \cite{Bel'kov79}\\
$0.2220\pm0.0128_{\rm stat}$
& Experiment on $K_{e4}$ decay & \cite{Batley-epjc10} \\
$\pm0.0050_{\rm syst}\!\pm\!0.0037_{\rm th}$ &  & \\
$0.220 \pm 0.005$ & ChPT + Roy's equations & \cite{CGL00,Colangelo01} \\
$0.220\pm 0.008$ & Dispersion relations & \cite{Garcia11} \\
  & and $K_{e4}$ data & \\
$0.26$ & NJL model (I)  & \cite{Volkov86} \\
$0.28$ & NJL model (II) & \cite{Ivan_Troi95} \\
\hline\hline
\end{tabular}}
\end{center}
\end{table}
In the analysis only of $\pi\pi$ scattering
and in the A-solution we reproduced with a high
accuracy the ChPT results \cite{CGL00,Colangelo01} including constraints
imposed by the Roy equations. On the other hand, the B-solution is
similar to the predictions of the chiral approaches based on the linear
realization of chiral symmetry (models of the
Nambu--Jona-Lasinio (NJL) type~\cite{Volkov86,Ivan_Troi95}).
Taking into account very precise
experiments at CERN performed by the NA48/2 \cite{Batley-epjc10}
and  the DIRAC \cite{DIRAC2011}  Collaborations, which confirmed the
ChPT prediction \cite{CGL00,Colangelo01}, one ought to prefer
the A-solution.

In summary, a structure of the Riemann surface of the $S$-matrix for coupled
processes must be included properly. To calculate resonance parameters, such
as masses and widths, one must use poles on those sheets where they are not
shifted (due to the channel coupling) in respect of the zeros on
sheet~I. In the two-channel case, the relevant poles are on sheets II or/and
IV depending on the resonance type.
Moreover, the combined analysis of coupled processes is needed as the
analysis of only $\pi\pi$ channel do not give correct values of resonance
parameters even if the Riemann surface structure is included. Finally, in order
to be concrete, the main scope of the paper is the scalar mesons, however, the
method and conclusions of paper can be applied to other wide resonances, e.g.
to vector mesons.

This work was supported in part by the
Heisenberg--Landau Program, the RFBR grant 10-02-00368-a, the Votruba--Blokhintsev
Program for Cooperation of Czech Republic with JINR, the GACR grant P203/12/2126,
the Grant Program of
Plenipotentiary of Slovak Republic at JINR, the Bogoliubov--Infeld Program
for Cooperation of Poland with JINR, the Polish Ministry of Science and Higher
Education (grant No N N202 101 368) and by the DFG under Contract No. LY 114/2-1.



\begin{thebibliography}{99}

\bibitem{PDG12}
J.Beringer {\it et al.} (PDG), Phys.\ Rev.\ D {\bf 86}, 010001 (2012).

\bibitem{Amsler04}
 C.~Amsler, N.A.~Tornqvist,
  Phys.\ Rept.\  {\bf 389}, 61 (2004).

\bibitem{Bugg04}
 D.V.~Bugg,
  Phys.\ Rept.\  {\bf 397}, 257 (2004).

\bibitem{Close02}
F.E.~Close, N.A.~Tornqvist,
  J.\ Phys.\ G G {\bf 28}, R249 (2002).
\textbf{28}, R249 (2002).

\bibitem{Klempt07}
 E.~Klempt, A.~Zaitsev,
  Phys.\ Rept.\  {\bf 454}, 1 (2007).

\bibitem{SBKN-PRD10}
Yu.S.~Surovtsev, P.~Byd\v{z}ovsk\'y, R.~Kami\'nski, M.~Nagy,
Phys.\ Rev.\ D {\bf 81}, 016001 (2010).

\bibitem{SBL-arXiv11}
Yu.S.~Surovtsev, P.~Byd\v{z}ovsk\'y, V.E.~Lyubovitskij,
Phys.\ Rev.\ D {\bf 85}, 036002 (2012).

\bibitem{KMS96}
D.~Krupa, V.A.~Meshcheryakov, Yu.S.~Surovtsev,
  Nuovo Cim.\ A {\bf 109}, 281 (1996).

\bibitem{SKN-epja02}
Yu.S.~Surovtsev, D.~Krupa, and M.~Nagy,
  Eur.\ Phys.\ J.\ A {\bf 15}, 409 (2002).

\bibitem{LeCou}
K.J. Le~Couteur, Proc. R. London, Ser. A \textbf{256}, 115
(1960); R.G.~Newton, J. Math. Phys. \textbf{2}, 188 (1961); M.~Kato,
Ann. Phys. \textbf{31}, 130 (1965).

\bibitem{SBKLN-1206}
Yu.~Surovtsev, P.~Byd\v{z}ovsk\'y, R.~Kami\'nski, V.~Lyubovitskij,
M.~Nagy,
arXiv: 1206.3438 [hep-ph].

\bibitem{Hya73}
B.~Hyams {\it et al.},
  Nucl.\ Phys.\ B {\bf 64}, 134 (1973)
  [AIP Conf.\ Proc.\  {\bf 13}, 206 (1973)];
Nucl.\ Phys.\ B {\bf 100}, 205 (1975).

\bibitem{Kamin02}
R.~Kami\'nski, L.~Le\'sniak, K.~Rybicki,
  Z.\ Phys.\ C {\bf 74}, 79 (1997);
  Eur.\ Phys.\ J.\ direct C {\bf 4}, 4 (2002).

 \bibitem{GarciaMKPRE-11}
R.~Garc{\'i}a-Mart{\'i}n,~R.~Kami{\'n}ski,~R.~Pel{\'a}ez,
J.~Ruiz de~Elvira,
Phys.\ Rev.\ Lett.\  {\bf 107}, 072001 (2011).

\bibitem{Rosselet77}
L.~Rosselet {\it et al.},
Phys.\ Rev.\ D {\bf 15}, 574 (1977).

\bibitem{Bel'kov79}
A.A.~Belkov {\it et al.},
JETP Lett.\  {\bf 29}, 597 (1979).

\bibitem{Achasov-Shest}
N.N.~Achasov, G.N.~Shestakov,
Phys.\ Rev.\ D {\bf 49}, 5779 (1994).

\bibitem{Batley-epjc10}
J.R.~Batley {\it et al.},
  Eur.\ Phys.\ J.\ C {\bf 70}, 635 (2010).

\bibitem{CGL00}
G.~Colangelo, J.~Gasser and H.~Leutwyler,
Phys.\ Lett.\ B {\bf 488}, 261 (2000).

\bibitem{Colangelo01}
G.~Colangelo, J.~Gasser, H.~Leutwyler,
Nucl.\ Phys.\ B {\bf 603}, 125 (2001);
B.~Ananthanarayan, G.~Colangelo, J.~Gasser, H.~Leutwyler,
Phys.\ Rept.\  {\bf 353}, 207 (2001).

\bibitem{Garcia11}
R.~Garc\'ia-Mart\'in, R.~Kami\'nski, J.R.~Pel\'aez, F.J.~Yndur\'ain,
Phys.\ Rev.\ D {\bf 83}, 074004 (2011).

\bibitem{Volkov86}
M.K.~Volkov,
Sov.\ J.\ Part.\ Nucl.\  {\bf 17}, 186 (1986).

\bibitem{Ivan_Troi95}
A.N.~Ivanov, N.I.~Troitskaya,
Nuovo Cim.\ A {\bf 108}, 555 (1995).

\bibitem{DIRAC2011}
B.~Adeva {\it et al.},
Phys.\ Lett.\ B {\bf 704}, 24 (2011).

\end{thebibliography}
\end{document}